# Double solution with chaos:
# Completion of de Broglie's nonlinear wave mechanics and its intrinsic unification with the causally extended relativity

A.P. KIRILYUK[*]

Institute of Metal Physics, Kiev, Ukraine 03142

ABSTRACT. As was shown previously (quant-ph/9902015), a simple system of interacting electromagnetic and gravitational protofields with generic parameters shows intrinsic instability with respect to unceasing cycles of physically real auto-squeeze (reduction) to a centre chosen by the system at random among many other, equally possible ones and the reverse extension, which form the causally probabilistic process of quantum beat observed as an elementary particle (like the electron). Here we show that the causally emerging wave-particle duality, space, and time lead to the well-known equations of special relativity and quantum mechanics thus providing their causal extension and intrinsic unification (also quant-ph/9911107). The relativistic inertial mass (energy) is universally defined as the temporal rate (frequency) of the chaotic quantum beat process(es) characterising their intensity. The same complex-dynamical processes and mass-energy account for the universal gravitation, since any reduction event in the electromagnetic protofield involves also the (directly unobservable) gravitational protofield increasing its tension and influencing the quantum-beat frequencies for other particles, which appears as gravitational interaction. This irreducibly complex dynamical mechanism of universal gravitation provides causal extension of general relativity intrinsically unified with causal quantum mechanics and special relativity, as well as physical origin and unification of all the four 'fundamental forces' (also gr-qc/9906077). The dynamic substantiation of the Dirac quantization rules is also obtained and used for first-principles derivation of the Dirac and Schrödinger equations which describe the same physically real, irreducibly complex (dynamically quantized and probabilistic) interaction processes within field-particles and their elementary systems (quant-ph/9511034-38). The classical, dynamically localised behaviour emerges as a higher level of complexity appearing as formation of elementary bound systems (like atoms), and the underlying picture of complex (irreducibly dualistic) dynamics agrees perfectly with recent observations of undulatory behaviour (beam diffraction, 'quantum condensates') of such classical systems with interaction. The derived intrinsic features of complex dynamics, such as dynamic uncertainty, quantization, and temporal irreversibility of the spatial structure formation, are reproduced at all higher levels of the naturally emerging hierarchy of complexity, which permits us to consistently understand the 'essentially quantum' behaviour as manifestation of the unreduced dynamic complexity of the world at its several lowest levels (physics/9806002).

---

[*]Address for correspondence: Post Box 115, Kiev-30, Ukraine 01030.
 E-mail: kiril@metfiz.freenet.kiev.ua

*RÉSUMÉ. Comme on a montré avant (quant-ph/9902015), un système simple des deux protochamps, électromagnétique et gravitationnel, interagissant entre eux possède, pour les valeurs génériques des paramètres, l'instabilité intrinsèque par rapport aux cycles répétitifs de l'auto-contraction (ou 'réduction') physiquement réelle vers un centre choisi par le système au hasard parmi plusieurs autres également possibles et l'extension inverse qui constituent le processus dynamiquement probabiliste de battement quantique observé comme une particule élémentaire (telle que l'électron). On montre ici que la dualité onde-corpuscule, l'espace, et le temps causals qui en émergent donnent les équations bien connues de la relativité restreinte et mécanique quantique qui acquièrent ainsi leur extension causale et l'unification intrinsèque (quant-ph/9911107). La masse-énergie inertielle relativiste est déduite universellement comme rapidité (fréquence) de changement temporel de processus chaotique(s) de battement quantique caractérisant leur intensité. Les mêmes processus dynamiquement complexes et la même masse-énergie résultent en gravitation universelle, car chaque événement de réduction du protochamp électromagnétique se reproduit dans le protochamp gravitationnel (directement inobservable) augmentant ainsi sa tension, ce qui influence la fréquence de battements quantiques d'autres particules et ainsi donne naissance à l'interaction gravitationnelle. Ce mécanisme constitue la base universelle, dynamiquement complexe de gravitation et l'extension causale de la relativité générale qui s'ensuit en l'unifiant en même temps avec les versions étendues de la mécanique quantique (ondulatoire) et la relativité restreinte (voir aussi gr-qc/9906077). L'origine physique et l'unification intrinsèque de toutes les quatre 'forces fondamentales' de la nature sont obtenues dans la même description. La justification causale et dynamique des règles de quantification de Dirac est également obtenu et utilisé pour la déduction consistante des équations de Dirac et Schrödinger qui décrivent les mêmes processus physiquement réels et essentiellement complexes (dynamiquement quantifiés et probabilistes) à l'intérieur des champ-particules et leurs systèmes élémentaires avec interaction (quant-ph/9511034-38). Le comportement classique, dynamiquement localisé émerge naturellement sur un niveau plus haut de complexité qui corresponde approximativement à la formation de systèmes liés élémentaires (comme les atomes), et la dynamique intérieure complexe (inévitablement dualiste) derrière cette interprétation se trouve en accord parfait avec les observations récentes des effets ondulatoires (diffraction de faisceau, 'condensates quantiques') dans ces systèmes classiques avec interaction. Les propriétés intrinsèques de la complexité dynamique obtenues, telles que l'incertitude dynamique, quantification, et l'irréversibilité temporelle de la formation de structures spatiales, sont reproduites sur tous les niveaux plus hauts de la hiérarchie de complexité naturellement émergent, ce qui nous permette de comprendre le comportement 'essentiellement quantique' comme manifestation de la complexité dynamique non-reduite du monde sur ses quelques niveaux les plus bas (physics/9806002).*

# 1. Quantum field mechanics as the unified causal extension of relativity and quantum mechanics

In previous works [1,2] it was shown that an elementary particle (exemplified by the electron) can be consistently described as a complex-dynamical process that arises in a system of two physically real interacting 'protofields' of electromagnetic (e/m) and gravitational nature, and consists in unceasing alternation of reduction (dynamically driven auto-squeeze) of a portion of the initially homogeneous fields to a very small volume and the reverse extension. The causal randomness (chaoticity) of this process of *quantum beat* originates from the *dynamic redundance* of the number of possible centres of reduction, each of them constituting an independent (locally complete) *realisation* of the system. Therefore the whole process can be described as unceasing series of spatially chaotic *quantum jumps* of the localised state of the elementary field-particle, or *virtual soliton*, from one unpredictably chosen centre of reduction to another. Now we are going to deduce an adequate description of the unreduced complex dynamics of such elementary field-particle at the level of observable quantities and shall call it here *quantum field mechanics*. The natural development of complex behaviour to its higher 'levels', including quantum measurement [3], quantum chaos [4], and classical behaviour [2], is also described within the proposed unifying concept of dynamic complexity.

Consider the elementary field-particle globally at rest. According to the previous analysis, it is represented by the unceasing, intrinsically random (probabilistic) process of quantum beat in the system of coupled e/m and gravitational protofields/media, equivalent to spatially chaotic wandering of the emerging virtual soliton. We can state now that this unceasing chaotic motion appears in observations as the universal property of *inertia* of the elementary particle (and eventually, of macroscopic bodies composed from such particles) measured by the field-particle *mass* which is equivalent to the *rest energy* appearing in the form of the chaotic wandering (realisation change) of the system. Indeed, it is clear that any attempt to change the dynamical state of the particle will meet an effective 'resistance' due to the *already existing* internal 'thermal' motion, which is somewhat similar to resistance to compression of a gas. Such definition of the property of inertial mass as an *emergent*, independent property is possible only due to the self-sufficient character of the quantum beat process and especially of its *purely dynamic* chaoticity directly related, in its turn, to the dynamic redundance/multivaluedness. In this way, we are able to extend the concept of the 'variable mass' of a particle in de Broglie's 'hidden thermodynamics' [5-7] because we can replace *additional* perturbations of the particle motion coming from the *external* 'hidden thermostat' ('subquantum medium') in the original de Broglie's picture by the *intrinsic* chaoticity *emerging* in the process of *a priory totally regular* protofield interaction.

In order to properly account for the unreduced complexity of the quantum beat dynamics, the mass-energy of the elementary particle thus introduced should be expressed through the unique and universal measure of that complexity naturally reflecting also the major property of discreteness of quantum behaviour. Since the latter always manifests itself through the quantum of *mechanical action*, or Planck's constant, $h$, we can suppose that it is the mechanical action, $\mathcal{A}$, which provides, within its extended understanding, such universal complexity measure. This means that each reduction-extension cycle of the *highly nonlinear, intrinsically unstable* quantum beat process corresponds to change of $h$ in the action value, $|\Delta\mathcal{A}| = h$, which evidently involves also a considerable extension of the notion of action itself with respect to the one from the essentially linear, unitary mechanics with its uniform evolution and smooth trajectory/path of the system motion. In addition, the function of action represents the only proper unification of the spatial and temporal aspects of dynamics emerging, as we have seen, within the quantum beat process [1,2]. In fact, action characterises the (spatial and temporal) distribution of probabilities of the dynamic reduction events described previously. Recalling also the canonical relation between energy, action, and time in the classical mechanics ($E = -\partial\mathcal{A}/\partial t$), we can summarise the obtained results within the following fundamental expression for the introduced rest energy (mass), $E_0$, representing it as the *temporal rate* (or intensity) of the fundamental quantum beat process:

$$E_0 = -\frac{\Delta\mathcal{A}}{\Delta t} = \frac{h}{\tau_0} = h\nu_0 \ , \tag{1}$$

where $-\Delta\mathcal{A} = h$ is the quantum of action-complexity corresponding *physically* to one cycle of the quantum beat (nonlinear reduction-extension of the coupled protofields or one 'quantum jump' of the virtual soliton), $\Delta t = \tau_0$ is the *emerging* 'quantum of time' equal to one period, $\tau_0$, of the same cycle, and $\nu_0 \equiv 1/\tau_0$ is the corresponding quantum beat frequency, forming the basis of the causal time concept [1,2] ($\nu_0 \sim 10^{20}$ Hz for the electron). Since according to the above definition energy is the measure of intensity, or inertia, of the internal chaotic dynamics of a particle, it should be proportional to particle mass, $E_0 = m_0 c^2$, where for the moment $c^2$ is simply a coefficient (we shall properly specify this relation later). This permits us to present the above expression for the rest energy, eq. (1), in an equivalent form,

$$m_0 c^2 = h\nu_0 \ , \tag{2}$$

which was semi-axiomatically introduced by de Broglie and further used in his original derivation of the wave properties of elementary particles and the famous expression for the 'de Broglie wavelength' [8]. We emphasize that in our version the simple relation of eq. (2) is causally *derived* from the *physically* based consideration of the underlying complex dynamics and represents just a compact expression of the causally complete picture of the highly nonlinear, dynamically redundant, causally probabilistic quantum beat process (in particular, the fundamental frequency $v_0$ represents something periodic, but actually much more involved than any linear, or even ordinary 'nonlinear' oscillation). Instead of justifying the left-hand side of eqs. (1), (2) by the *abstract* mass-energy equivalence from the canonical relativity which is *formally* deduced itself from the *postulates* about the *abstract laws* of nature, *without* any *physically complete* understanding of *what* mass, energy, and time *actually are*, we introduce energy (and the *causally* equivalent mass) as the *emerging* temporal rate of the *causally obtained* process of *chaotic* realisation change (which is completely specified as the *physical* reduction-extension of the coupled primal protofields). This is the beginning of the natural appearance, within our description, of the *causal*, physically based relativity which is *intrinsically unified* with the quantized wave dynamics of the elementary field-particle and is simply another expression of the *same*, universal *dynamic complexity* as that underlying the quantum behaviour of the *same* system of coupled protofields.

If now the isolation of the elementary field-particle is violated and it is subjected to external influences (from other particles), then the intensity of the quantum beat, and thus particle energy, can change. We can rigorously define the *state of rest* of the elementary field-particle (and eventually of a macroscopic body consisting from such particles) as the state with *minimum* complexity-energy. This minimum exists, since the quantum beat energy is always positive and finite. For the elementary field-particle it is realised as totally irregular spatial distribution of the reduction centres, which corresponds to the absolutely homogeneous distribution of realisation probabilities causally deduced previously [1,2] within the generalised Born's rule or simply as a result of 'equal rights' for existence of apparently equivalent realisations. Therefore in the case of single particle the energy minimum is unique, and any change would correspond to violation of its internal 'maximal irregularity' (homogeneity of the probability distribution).[1]

---

[1] This result can be generalised to any higher level of complex behaviour, since if the introduced state of rest is not unique, it means that the system contains additional redundance, leading to corresponding chaotic transitions between those multiple 'states of rest', and one should actually consider this additional chaoticity at the new, thus emerging level of complexity possessing now the unique state of rest with maximal irregularity.

It is easy to make the straightforward next step and introduce an equally rigorous and universal definition of a *state of motion* as a one characterised by the complexity-energy *exceeding* the minimum value of the state of rest for the given system. Naturally, there can be many such higher values of energy and many respective states of motion. Since the state of rest has a pathologically unique, totally uniform spatial structure, it is clear that any state of motion of the field-particle will be characterised by an *inhomogeneous* spatial distribution of reduction probabilities, corresponding to a dynamical *tendency* in the thus emerging *global* motion of the field-particle. Correspondingly, any state of motion of a more complex object is characterised by a more inhomogeneous distribution of realisation probabilities than that for its state of rest. Whereas the centre of *each* next reduction event is 'chosen' by the system always in a *purely* probabilistic fashion, there is now more 'order in chaos', and the emerging more inhomogeneous *structure* in the reduction *probability distribution* manifests itself as *observable spatial degrees of freedom* of the appearing 'level of complexity', that is the particle/object 'displacement' as such. In terms of the proposed mathematical measures of complexity, this means that our action-complexity $\mathcal{A}$ for the field-particle in the state of motion acquires a (regular) *spatial dependence* (structure), whereas in the state of rest it depends only on time. Therefore the partial time derivative of action of the moving field-particle, always defining its energy, is now different from the total time derivative:

$$\frac{d\mathcal{A}}{dt} = \frac{\partial \mathcal{A}}{\partial t} + \frac{\partial \mathcal{A}}{\partial x}\frac{dx}{dt} = -E + pv , \tag{3}$$

where the *momentum*, $p$, that characterises the emerging spatial order in the probability distribution of the moving field-particle and its *velocity*, $v$, are introduced in accord with the canonical relations:

$$p = \frac{\partial \mathcal{A}}{\partial x}, \quad v = \frac{\partial x}{\partial t} .$$

Now we should take into account the natural discreteness (dynamic quantization) of the quantum beat process meaning, in particular, that the spatial field-particle structure can also emerge only as discrete elements with spatial *dimension* (size) $\lambda$ determined always by the same quantum of action, $h$:

$$p = \frac{\Delta \mathcal{A}}{\Delta x}\bigg|_{t=\text{const}} = \frac{\Delta \mathcal{A}}{\lambda} = \frac{h}{\lambda} , \tag{4}$$

$$v = \frac{\Delta x}{\Delta t} \equiv \frac{\lambda}{T} . \tag{5}$$

where $\lambda \equiv (\Delta x)|_{t \,=\, \text{const}}$ is the *emerging* 'quantum of space', a minimum directly measurable (regular) space inhomogeneity characterising the elementary quantum field with complexity-energy $E$ ($> E_0$) and resulting from its global displacement (motion), $\Delta t = T$ is the 'total' period of nonlinear quantum beat of the field in the state of motion with complexity-energy $E$ ($N = 1/T$ is the corresponding quantum-beat frequency), $\Delta x = \Lambda$ is the 'total' quantum of space, while $\tau = (\Delta t)|_{x \,=\, \text{const}}$ is the quantum-beat oscillation period measured at a fixed space point (so that $E = h/\tau$). We can therefore rewrite eq. (3) in the following form, specified for the quantum beat dynamics of the moving field-particle:

$$E = -\frac{\Delta \mathcal{A}}{\Delta t} + \frac{\Delta \mathcal{A}}{\lambda} \frac{\Delta x}{\Delta t} = \frac{h}{T} + \frac{h}{\lambda} v = hN + pv , \qquad (6)$$

where

$$E = -\frac{\Delta \mathcal{A}}{\Delta t}\bigg|_{x=\text{const}} = \frac{h}{\tau} = h\nu . \qquad (7)$$

This relation replaces that of eqs. (1), (2) for the same field-particle at rest, and includes the causal definitions of the emergent space, time, momentum, and energy, applicable in the general case.

Note that the characteristic temporal intervals for the moving particle ($\tau$ and $T$) are both different from the single time scale $\tau_0$ for the particle at rest because the modified (and non-unique) intensity, and thus temporal rate, of the quantum beat process constitutes the physically transparent essence of the global motion: in order to advance as a whole, the particle should *relatively* (with respect to the state of rest) intensify (accelerate) its reduction-extension cycles 'in the direction of motion' at the expense of those in other directions, which reveals already the fundamental source of the causal 'relativity of time' (it will be further specified below). We emphasize the fundamental and objective character of our definition of motion and rest not depending upon any externally inserted, intuitive 'observations of displacement', etc. (though being in accord with them), which is possible due to the universal guiding role of the unreduced, holistic complexity underlying the coupled protofield dynamics that always *autonomously* performs a (generalised) 'observation', or 'measurement', *on itself*.

We see also that the appearance of global motion in that dynamics is naturally involved with emergence of an elementary spatial structure of the moving field-particle, in the form of an *average* regular *tendency* within the generally chaotic wave field. It is not difficult to understand that this structure is none other than the causally extended, realistic 'de Broglie wave of the (moving) particle', whereas $\lambda \equiv \lambda_B$ is the 'de Broglie wavelength', and we shall continue to confirm and specify this conclusion below. The physical sense of the energy partition of eq. (6) becomes clear: the second term, $pv = hv/\lambda$, corresponds to the regular, global motion tendency, or structure, whereas the first term, $-(d\mathcal{A}/dt) = h/T$, represents the purely irregular, 'thermal' wandering of the virtual soliton 'around' the average tendency/structure. This dynamically based interpretation confirms and completes the corresponding 'thermodynamical' considerations of de Broglie. Note that the internal motions within the moving field-particle thus described can be figuratively presented in an intuitively transparent image of a complex-dynamical 'caterpillar motion', where an externally smooth particle motion is obtained in reality as a result of many superimposed protofield reductions to a (physical, emerging) point accompanied by 'pulling' of the extended wave field 'body' to that point, etc. [2].

Further refinement of mathematical description of the quantum beat dynamics for the moving field-particle, eq. (6), should express a relation between the temporal ($E$) and spatial ($p$) rates of the complex-dynamical structure formation process called, in accord with the corresponding relation in the canonical science, *dispersion relation*. The holistic character of the quantum beat dynamics shows that such a relation should exist in a well-defined form and can be rigorously deduced: the regular global structure of the wave field is formed by the same reduction-extension cycles as those determining the general quantum beat intensity. The dispersion relation we are looking for can be considered as the causally completed version of the 'phase accord theorem' introduced by de Broglie in his original substantiation of existence of a wave associated with a particle [8]. While the wave existence is *postulated* and its physical origin remains unclear within this original formulation, de Broglie shows that if the 'internal oscillations' of the particle, also *postulated* by the heuristically composed analogue of eq. (2), remain always in phase with those of the wave (i. e. physically the wave performs the stationary transport of, 'pilots', or 'develops', the particle-oscillation) then such 'compound object' moves in accord with the relativistic transformations of time and mass (also *formally postulated* in the standard theory) and the length of the wave is given by the (now) canonical expression for the 'de Broglie wavelength'. The quantum beat dynamics involves a self-consistently unified, causal extension of the participating entities and their 'phase accord': the localised 'particle' (virtual soliton) moves always 'in phase' with the extended wave ('intermediate realisation' [1,2]) simply because it is permanently *transformed* into that wave and back, in course of its 'oscillations' (i. e. succesive reductions and extensions). In other words, the internal 'particle oscillations' and the 'piloting' wave represent two *dynamically* related, alternating and coexisting *nonlinear* aspects of *one and the same* quantum beat process. Another essential extension of the original phase-accord idea is in the fact that the regular, averaged structure of the entire wave field, forming the causal analogue of the 'transporting wave', represents only a part of the whole dynamics, and the localised virtual soliton, or 'particle', makes many pronounced *deviations* from that regular tendency 'taking' with it the corresponding extended wave structures that form a causally random, irregular part of the wave field, always preserving nonetheless the internal 'phase accord' between 'particle' and 'wave'. It is clear, however, that only the averaged regular structure of the wave field, the causally completed version of the 'de Broglie wave', can be directly observed in experiments as a wave, while the stochastic, purely random components of the field-particle dynamics appear only 'all together', in the form of the particle (rest) mass.

Consider a portion of that regular undulatory spatial structure of the wave field in the state of global motion incorporating $n$ (de Broglian) wavelengths, $x_0 = n\lambda$. The *temporal*, oscillatory aspect of the quantum beat process determining the (total) energy covers the *same* measured distance *in the direction of motion* remaining in the described involvement with the wave, but performs simultaneously an intensive irregular, 'sideways' deviations (wandering). Since the velocity of the virtual soliton propagation along the actual irregular path should always be equal to the velocity of light, $c$ (it is the velocity of perturbation propagation in the directly observable e/m medium), it becomes clear that the energy-bearing oscillation of the quantum beat performs $n' = n(c/v)$ cycles within the same portion of the regular displacement with velocity $v$. It is *physically* clear why always $n' > n$ and thus $v < c$: the difference between $n'$ and $n$ just accounts for the 'hidden', purely irregular part of the quantum beat process. The measured advance of the oscillation (temporal aspect), $x_1 = n'\tau c = n(c^2/v)\tau$, should be equal to the observed displacement of the wave (spatial aspect), $x_0 = n\lambda$, within the unique quantum beat process: $x_0 = x_1$, or $\lambda = V_{ph}\tau$, where $V_{ph} = c^2/v$ is the fictitious superluminal 'phase velocity' introduced by de Broglie within the original phase accord conjecture [8]. Rewriting the obtained relation between $\lambda$ and $\tau$ as

$$\frac{1}{\lambda} = \frac{1}{\tau}\frac{v}{c^2},$$

multiplying it by $h$, and using the definitions for momentum and energy, eqs. (4), (7), we finally obtain the desired dispersion relation between spatial (regular) and temporal (total, regular and irregular) aspects of complex field-particle dynamics:

$$p = E\frac{v}{c^2} = mv, \qquad (8)$$

where $m \equiv E/c^2$, now by rigorously substantiated definition. We have thus causally deduced the famous 'relativistic' dispersion relation from the detailed analysis of the unreduced *complex dynamics* actually underlying any *externally* 'uniform' motion (whereas it is derived from *formal postulates* in the canonical theory, see e. g. [9]) and simultaneously specified the law of proportionality and the coefficient, $c^2$, in the still more famous energy-mass relation, already physically substantiated above and also only mechanistically 'guessed' in the standard relativity.

We can see now the origin of the characteristic inconsistency around the superluminal de Broglian 'wave of phase' [8], being both physically *unreal* and indispensable for the proposed *realistic* description: if one does not *explicitly* take into account the *really* existing, *intrinsic* irregular part of the unique, *inseparable* quantum beat dynamics, then the corresponding part of energy is 'forced' to reappear in a reduced form of *regular* motion, inevitably becoming 'superluminal' because of its 'excessive' character. The issue found by de Broglie and transforming the ambiguous 'wave of phase', propagating with the velocity $V_{ph} = c^2/v$, into a real wave transport of the particle by the linear *wave packet* moving with the proper *group velocity* $v$, cannot really solve the problem, since the purely *linear* transformation between the phase and group wave propagation shows fundamental divergence with the necessarily and essentially *nonlinear* character of the wave mechanics (in particular, the linear wave packets quickly spread, etc.). The realistic undulatory aspect of the moving elementary particle should indeed be obtained from *many* participating, slightly *differing* components, but in an adequate, essentially nonlinear description they should necessarily (strongly) *interact* with each other, which leads, as we show within the substantiation of the quantum field mechanics [1,2], to the basically *unstable* character of the resulting quantum beat dynamics, where the nonlinear component interaction takes the form of their ruthless *competition* in which only those corresponding to the current tendency (reduction to a randomly selected centre or the opposite extension) can survive and form both the localised 'particle' and its 'wave transport'. Only the unreduced involvement of the largely hidden, but really existing complex-dynamical processes can consistently explain the externally simple 'relativistic' and 'quantum' relations between mass, energy, and momentum, eqs. (2), (8), and it is the insertion of the deduced dispersion relation, eq. (7), into the emerging momentum definition, eq. (4), that finally closes the causally complete substantiation of the de Broglie expression for the wavelength:

$$\lambda \equiv \lambda_B = \frac{h}{mv} \quad . \tag{9}$$

We emphasize the crucial importance of the obtained dispersion relation for the complete understanding of the meaning of eq. (9). It is the deceptively familiar, 'classical' and almost 'trivial' relation between the particle momentum, mass, and velocity, $p = mv$, which bounds together the wave-like, corpuscular (classical) and relativistic properties of the elementary field-particle into the intrinsically inseparable (within the quantum field mechanics) and most 'mysterious' (within the canonical 'interpretations') mixture summarised by the fundamental de Broglie equation. Note also that the same dispersion relation is actually equivalent to the main dynamical laws of the classical mechanics (Newton's laws and their relativistic extension) which are obtained in their canonical form by taking the time derivative of eq. (8). We see now that these laws can be causally derived, instead of being postulated in the canonical approach, if we take into account the internal complex (chaotic) dynamics of any motion, including all the externally 'uniform' and 'rectilinear' cases (the mechanism of classical behaviour emergence is described below).

The emerging causal unification of the (extended) relativity and quantum mechanics within the quantum field mechanics does not stop there, and we continue to specify it by inserting the causally derived dispersion relation, eq. (8), into the basic energy partition of eq. (6) and using the complex-dynamical energy definition, eq. (7), which gives:

$$\tau = T\left(1 - \frac{v^2}{c^2}\right). \qquad (10)$$

This relation provides further refinement of the *physical* explanation of the canonical 'relativistic time retardation' inevitably inseparable from the causally complete understanding of the entity of time itself. Since time naturally *emerges together* with structure-forming reduction-extension events within the quantum-beat *motion*, the increased intensity (rate) of this process for the (relatively) moving field-particle(s) is realised as a relative *decrease* of the elementary quantum beat period, $\tau$, which results in *relative* retardation of *any other* process from a higher level of complexity within the moving group of particles (a physical 'body'), *irrespective* of their detailed dynamics (their duration with respect to the state of rest is proportional to $T$). A related equivalent explanation involves 'transport effects' accompanying the appearing undulatory structure and determined by the difference between the total and partial time derivatives of the action-complexity (see eq. (3)): they lead to a relative *increase* of the *local* 'total-change' (irregular-motion) period $T$ which just determines the relative local 'lifetime' of the elementary, or any higher-complexity, motion cycles. In other words, the causal time within the moving system, *with respect* to the state of rest, goes as $nT$ (with permanently growing integer $n$), and the relative duration of any given process equals to $n_0 T$, where $n_0$ depends only on the process type, but not on its global state of motion. In order to obtain the causal time retardation effect in the explicit form, we should now express the quantum beat periods $\tau$ and $T$ through the reference period in the state of rest, $\tau_0$. It is not difficult to see [2] that the corresponding frequencies, $\nu$, $N$, and $\nu_0$ are related by the following equation:

$$N\nu = (\nu_0)^2 , \tag{11a}$$

which gives, for the periods

$$T\tau = (\tau_0)^2 . \tag{11b}$$

These relations can be reduced to a physically transparent law of 'conservation of the total number of reduction events' which is a manifestation of the universal complexity conservation law [2]. In our case, it stems simply from the fact that the driving electro-gravitational coupling remains unchanged for any dynamical state of the field-particle, and therefore the total number of reduction events (per unit of time), including both the regular tendency and irregular deviations, should be the same in the states of rest and (uniform) motion (respectively the right- and left-hand sides of eq. (11a)). Using eqs. (11) in conjunction with eq. (10), we get the canonical expression of the time retardation effect, now causally extended by the underlying physical picture of quantum beat dynamics:

$$T = \frac{\tau_0}{\sqrt{1 - \frac{v^2}{c^2}}} \quad \text{or} \quad N = \nu_0 \sqrt{1 - \frac{v^2}{c^2}} , \tag{12a}$$

$$\tau = \tau_0 \sqrt{1 - \frac{v^2}{c^2}} \quad \text{or} \quad \nu = \frac{\nu_0}{\sqrt{1 - \frac{v^2}{c^2}}} . \tag{12b}$$

We can now combine in one expression the complex-dynamical partition of the total energy into the regular transport and irregular ('thermal') wandering, eq. (6), with the causally deduced dispersion relation, eqs. (8), (9), and time (frequency) relation to dynamics, eqs. (12):

$$E = h\nu_0 \sqrt{1 - \frac{v^2}{c^2}} + \frac{h}{\lambda_B} v = h\nu_0 \sqrt{1 - \frac{v^2}{c^2}} + h\nu_B = h\nu_0 \sqrt{1 - \frac{v^2}{c^2}} + \frac{m_0 v^2}{\sqrt{1 - \frac{v^2}{c^2}}} ,$$

$$\tag{13}$$

where $h\nu_0 = m_0 c^2$ (eq. (2)) and we introduce *de Broglie frequency*, $\nu_B$, defined as

$$\nu_B = \frac{v}{\lambda_B} = \frac{pv}{h} = \frac{\nu_{B0}}{\sqrt{1 - \frac{v^2}{c^2}}} = \frac{v^2}{c^2} \nu , \; \nu_{B0} = \frac{m_0 v^2}{h} = \nu_0 \frac{v^2}{c^2} = \frac{v}{\lambda_{B0}} , \; \lambda_{B0} = \frac{h}{m_0 v} . \tag{14}$$

The summarised expression of the complex dynamics of a moving field-particle, eqs. (13), (14), including the 'ordinary' type of relation between the causal de Broglie wavelength, frequency, and wave/particle velocity, $\lambda_B v_B = v$, clearly demonstrates the physical reality of this wave and demystifies its origin. In particular, $v_B$ is the *average* frequency of the dynamically quantized propagation of the 'travelling' de Broglie wave that advances to $\lambda_B$ during each indivisible and irregularly occurring step. In this respect the averaged, de Broglian wave field of the moving field-particle resembles, due to the internal nonlinearity, both travelling and standing wave, the latter having spatially fixed nodal planes. This averaged, global tendency, represented by the second summand in eqs. (13), could not exist without purely irregular deviations from it ('thermal motion'), represented by the first term in eqs. (13), and their ratio, $R$, equal to the ratio of the corresponding frequencies, shows the *relative* number of probabilistic quantum jumps falling within the regular tendency and taken with respect to the number of purely irregular events:

$$E = m_0 c^2 \sqrt{1 - \frac{v^2}{c^2}} [1 + R] \, , \, R \equiv R(v) = \frac{v^2/c^2}{1 - \frac{v^2}{c^2}} = \frac{\beta^2}{1 - \beta^2} \, , \, \beta = \frac{v}{c} \, . \quad (13')$$

It is clear that $\alpha_1 = \beta^2 = v^2/c^2$ and $\alpha_2 = 1 - \alpha_1 = 1 - \beta^2 = 1 - v^2/c^2$ are the dynamically defined probabilities that an elementary reduction-extension event falls within, respectively, the regular (averaged, global) and purely irregular ('thermal') tendency of the complex quantum-beat dynamics, while $R = \alpha_1/\alpha_2 = \alpha_1/(1 - \alpha_1)$ (cf. the general definition of such 'compound realisation' probabilities, $\alpha_j$, in [1,2]). We see that a slow, 'nonrelativistic' propagation of the field-particle ($v \ll c$, $\beta \ll 1$) is characterised by a very weak 'order in chaos', $R, \alpha_1 \ll 1$, $\alpha_2 \cong 1$, which means that the distribution of realisation (reduction) probabilities is only slightly inhomogeneous and only a small portion of the probabilistic 'quantum jumps' of the virtual soliton and the related wave-field structures falls within the regular tendency forming de Broglie wave. The structure of the probabilistic wave field becomes pronouncedly ordered, $R \sim 1$, $\alpha_1, \alpha_2 < 1$, only at moderate relativistic velocities of the global displacement ($v < c$, $\beta < 1$). And finally, it is almost totally ordered, $R \gg 1$, $\alpha_1 \cong 1$, $\alpha_2 \ll 1$, at 'ultra-relativistic' velocities ($v \cong c$, $1 - \beta \ll 1$).

Thus a causally complete, dynamical sense of 'relativistic' motion and velocity is revealed within the unreduced picture of complex dynamics of the elementary field-particle. The 'relativism' of dynamics is an explicit manifestation of its complexity, fundamentally inseparable from 'quantum' manifestations of the *same* dynamic complexity (quantized spatial and probabilistic temporal structure), even though the two types of manifestations *seem* not to be directly related both in practical observations, and within the canonical, unitary versions of quantum mechanics and relativity. The increasingly relativistic character of dynamics with the growing global motion velocity corresponds to growing dynamic complexity, measured by such related quantities as $E$, $p$, $R$, and $\beta$ Physically this change corresponds to increasing order within the causally chaotic dynamics, *always* preserving, however, its intrinsically probabilistic character at *every single step* (quantum jump of the virtual soliton). This latter feature explains why any dynamically complex, *massive* particle-process cannot not only exceed, but even attain the velocity of light: the condition $v = c$ ($\beta = 1$) would correspond to the totally regular, zero-complexity dynamics, where the field-particle cannot have any possibility for *irregular deviations* from the global tendency corresponding to the property of mass. Only massless, *totally regular* perturbations of the e/m protofield (like photons [2]) propagate with the speed of light. We can also *physically* understand now why exactly the global, regular motion of a massive field-particle (and thus of any 'ordinary', structure-forming matter) necessarily involves the 'relativistic' increase of *inertial* mass, i. e. why *any* energy possesses the property of inertia: it is because *any* part of the total complex process of motion, including the regular *in average* global tendency, occurs through an intrinsically *random* choice of reduction centres giving rise to inertia, according to the above causal interpretation of the latter. That is why, as follows from eqs. (13´),

$$m = \frac{E}{c^2} = m_0 \sqrt{1 - \frac{v^2}{c^2}} \left[ 1 + \frac{v^2/c^2}{1 - \frac{v^2}{c^2}} \right] = \frac{m_0}{\sqrt{1 - \frac{v^2}{c^2}}} \ . \tag{15}$$

The energy-mass relation itself can be interpreted in terms of the complex-dynamical motion partition into global (regular) and local (irregular) parts: the famous $mc^2$ simply corresponds to the global motion energy, $mv^2$ (the last term of eqs. (13)), at $v = c$, which means that if all the probabilistic jumps within the complex-dynamical process with the total mass $m$ (each of them proceeding indeed at velocity $c$) could be 'aligned' into one totally regular tendency, then we would obtain only the regular-tendency part of energy, $E = mc^2$, while the irregular part would be absent (something like this can really happen during an annihilation process).[2] It is clear also why various potential motions contribute to the total mass-energy-complexity, in the form of 'potential energy of interaction': they represent a hidden stock of yet unrealised, 'future' (but real) complexity of the quantum beat dynamics (and in the general case also other, higher-level complex processes) which will evolve into the corresponding change of the explicitly appearing temporal (quantum-beat frequency) and spatial (de Broglie wavelength) forms of complexity [2].

---

[2]Similar considerations show that a characteristic spatial structure size of the irregular part of the particle wave field — and actually the minimum realisable spatial dimension (for the given field-particle) — corresponds to momentum $m_0 c$ and thus equals to the Compton wavelength, $\lambda_C = h/m_0 c$. This smallest wave-field inhomogeneity coincides with the length of the elementary quantum jump of the virtual soliton performing its causally random wandering within the field-particle at rest [2].

We can conclude that all the canonical effects and relations of 'special relativity', including the 'relativity of time' (and space), can be considered, in their *causally complete* version, as particular observable manifestations and quantified expressions of the extended 'hidden thermodynamics', the 'double solution with chaos', thus giving a clear response to accusations in the 'absence of experimental manifestations' of the results of de Broglie's approach: the latter endows the 'well-known' relativistic relations with the physically real, logically rigorous (truly first-principles) substantiation that was absent in the canonical, Einsteinian interpretation and could *not* be obtained in principle within *any single-valued* approach of the canonical science. In this way the double solution with chaos provides also the natural unification of the first and second laws of thermodynamics (energy conservation and entropy increase, respectively), or homogeneity and irreversibility of the related causal time, constituting an aspect of the ultimate unification of all observed entities, formal laws and principles of the canonical science within the causally substantiated, universal law of conservation (symmetry) of complexity [2] (see also below).

One may additionally inquire about the sense of the 'classical' relativistic expressions for energy, eqs. (13)-(15), obtained for the essentially quantum, undulatory behaviour of the elementary field-particle. As we have seen above, they refer rather to the chaotic wandering of the localised, particle-like, though always transient, state of the virtual soliton. We can say that the real 'trajectory' of this 'particle' is very irregular, it literally 'changes at every its point', due to quantum jumps of the virtual soliton (where 'points' are small, but finite and physically well defined [1,2]). This kind of motion of a localised complex-dynamical object-realisation can be represented in the form of the 'generalised Lagrange formalism' [2], being simply another aspect of the above picture summarised in eqs. (13) and applicable also at higher levels of complexity.

By analogy to classical mechanics and in accord with the main energy-partition relation of eq. (3), the generalised Lagrangian, $L$, is defined as the discrete analogue of the *total* time derivative of action-complexity $\mathcal{A}$:

$$L = \frac{\Delta \mathcal{A}}{\Delta t} = pv - E \ . \tag{16}$$

Recalling the detailed expressions and physical sense of the 'de Broglian' energy partition, eqs. (6), (13), we see that $L$ is simply the 'thermal', purely irregular part of the total energy taken with the opposite sign and for the isolated field-particle is obtained as

$$L = -hN = -m_0 c^2 \sqrt{1 - \frac{v^2}{c^2}} \ . \qquad (17)$$

We find again a remarkable coincidence with the corresponding expression from the canonical relativity [9], where it is mechanistically 'guessed' by purely formal considerations and then actually postulated, whereas in the quantum field mechanics we not only causally deduce the expression of eq. (17), but reveal the physically complete picture of chaotic dynamics behind it. In particular, we obtain the universal *physical* interpretation of the Lagrangian, valid in principle for any dynamical system (cf. below). Namely, the function of Lagrangian used for formal description in various fields of the canonical theoretical physics represents the purely irregular component of energy of a considered (necessarily internally chaotic) motion, normally taken with the negative sign. The total energy generally includes, in addition to this 'thermal' part (-*L*), also the regular (global) motion energy, *pv*.

The Lagrangian definition, eq. (16), can be written in integral form,

$$-hn = \sum_{j=N+1}^{j=N+n} L_j T_j \ , \qquad (18)$$

where the summation is performed over a portion of the chaotic virtual-soliton 'trajectory' represented by its consecutive positions/realisations, $T_j$ is the full time increment at the *j*-th step (quantum jump), and $L_j$ is the corresponding Lagrangian value. Each complex-dynamical, 'quantum' jump of the virtual soliton-particle corresponds to addition of a negative action quantum, -*h*, to both sides of eq. (18). It is easy to see that this expression is the complex-dynamical generalisation of the classical 'action integral',

$$S_{\text{class}} = \int_{t_1}^{t_2} L \, dt \quad . \qquad (19)$$

*Formal* variation of trajectorial point coordinates and velocities, within the canonical Hamilton-Lagrange approach, leads to variation of *L* and $S_{\text{class}}$ from eq. (19) corresponding to many *fictitious*, possible trajectories, and the canonical 'variational principle' ('principle of least action' in mechanics) states that in reality the system takes only one of them, and namely the one for which the value of $S_{\text{class}}$ is minimal (which gives the main differential equations of the canonical dynamics). Correspondingly, being taken over the whole accessible domain of the field-particle motion (i. e. over all system realisations), the unreduced, complex-dynamical result of eq. (18) leads to an important extension of the canonical 'variational principle' stating now that instead of formal variations and fictitious trajectories, a complex-dynamical system, represented here by the quantum-beat dynamics of the elementary field-particle, performs *real* deviations (causally random jumps) along a *real* chaotic trajectory, including the points of former 'fictitious' trajectories occupied with the well-specified, dynamically determined probabilities. Any observed 'unique' trajectory can result from this real chaotic wandering of the system as an averaged, always partially smeared trajectory which is sufficiently well defined only in cases of localised, 'classical' type of behaviour (we describe below how exactly it emerges). The canonical 'principle of least action' is replaced now by the universal *complexity conservation law* [2]. Such extension of the least action principle is valid for arbitrary system at any level of complexity [2] and has been predicted by Louis de Broglie, starting from the hidden thermodynamics idea [7]. It is physically clear, from the above causal interpretation of the Lagrangian, why it is precisely this, irregular-motion part of the total energy that enters the generalised principle of least action.

Moreover, the particularly explicit appearance of quantized chaotic jumps of the complex-dynamical system in the extended Hamilton-Lagrange formulation of eq. (18) directly emphasises their role as the specific mechanism of strong *nonunitarity*, i. e. unpredictability and non-uniformity, of real system evolution. It is easy to see that because of such universal jump-like behaviour, the evolution of a generic (quantum or classical) system cannot be presented, in any sense, by exponential functions in the canonical expressions for 'evolution operators', 'Feynman path integrals', etc. In particular, the canonical path integrals, also referring to 'fictitious' paths, are actually replaced by the same universal expression for the action-complexity, eq. (18), or its 'undulatory' version (see below) [2]. Using eqs. (16), (18), (19), one can also easily obtain the causally extended and universally applicable versions of 'Bohr's quantization rules' and 'Heisenberg uncertainty relations' which provide an additional completion of the quantum field mechanics and full understanding of the physical origin of the corresponding 'mysterious' features from the ordinary quantum mechanics (see [2] for more detail).

If we want to obtain a really adequate mathematical description of the intrinsic wave-particle duality of the quantum beat process, then the above trajectorial, localised type of formalism should be completed by a naturally obtained wave-like, delocalised type of formal presentation accounting for the causally probabilistic wave field of the field-particle in the extended phase of the 'intermediate' realisation. This partially ordered, dynamically maintained structure of the e/m protofield interacting with gravitational matrix-protofield [1,2] is described by the *physically based* 'wavefunction', $\Psi$, which arises from the initial 'state-function' of the interacting protofields [1] and is the causal extension of the canonical wavefunction introduced by Erwin Schrödinger. The essential difference between the two is due to the complex, dynamically chaotic physical origin of the extended version, which naturally resolves the basic difficulties of the canonical description around *imposed* probabilistic 'interpretation' of $\Psi$ and permits us to preserve the *direct* relation of the mathematical function $\Psi$ with the characteristics of the physically real, *dynamically probabilistic* wave field of the quantum beat process and at the same time causally justify 'Born's probability rule' interpreting $|\Psi|^2$ as the particle emergence probability (see refs. [1,2] for its detailed explanation in the general case). Simultaneously we resolve the related problem of the purely formal 'configurational space' providing the 'independent variable' on which the canonical wavefunction depends: the causal wavefunction is defined on the *physically real* space of *dynamically emerging* realisations, the latter providing the causal version of 'configurations' which are reduced, in the simplest case, to the physical 'space points' (reduction centres) described previously [1,2]. We see also that the well-specified *nonlinearity* of the particular system of real physical (proto)fields, originating simply from their *unreduced interaction* and leading to the dynamic redundance, permits us to replace a rather sophisticated and somewhat ambiguous combination of *several* components of an additional, specially introduced 'quantum field' in the original version of the double solution [10-12] by the *single* causally obtained wavefunction representing a *real* dynamical state of the participating *physical* fields. This state can be presented as a superposition of the regular, averaged wave-field structured with the de Broglie wavelength and the irregular wave field with a characteristic structure of the size of the Compton wavelength. Note that the causal wavefunction is a version of the 'state-function' introduced previously [1,2] for description of the interacting protofield dynamics and considered here at a somewhat higher 'sublevel' of complexity, where all the details of the protofield configuration cannot be observed, but only their dynamically 'coarse-grained', quantized structure.

In order to obtain the delocalised, wavefunctional description in the dual unity with the above localised, corpuscular description, we can suppose that the corresponding quantities, action-complexity $\mathcal{A}$ and wavefunction $\Psi$, are entangled in a single quantity of 'wave action', $\mathcal{A}_\Psi$, representing the total dynamic complexity of the quantum beat process, in all its aspects:

$$\mathcal{A}_\Psi = \mathcal{A}\Psi \ . \tag{20}$$

The wave action is close to the action-complexity of the virtual soliton motion, but explicitly takes into account the existence of the dynamically related extended phase (realisation). The physical basis of this relation between $\mathcal{A}_\Psi$, $\mathcal{A}$, and $\Psi$ becomes more transparent if we consider its change during one cycle of reduction-extension. The change of $\mathcal{A}_\Psi$ should be equal to zero, since the total complexity is permanent according to the complexity conservation law (it simply expresses the permanence of quantum beat in the isolated system of the coupled protofields). Indeed, if the cycle starts from the extended state, then it should end up with the *same* state, since there is *only one* intermediate realisation [1,2] 'bonding together' all the 'regular', localised realisations of the virtual soliton. Therefore

$$\Delta \mathcal{A}_\Psi = \mathcal{A}\Delta\Psi + \Psi\Delta\mathcal{A} = 0 , \qquad (21a)$$

or

$$\Delta\mathcal{A} = -h \frac{\Delta\Psi}{\Psi} , \qquad (21b)$$

since the characteristic value of action $\mathcal{A}$ during the cycle is $h$. The latter is additionally multiplied by a numerical constant, $i/2\pi$, which does not change the physical sense of the above duality expression and accounts for the difference between wave and corpuscular states in the wave presentation by complex numbers:

$$\Delta\mathcal{A} = -i\hbar \frac{\Delta\Psi}{\Psi} , \qquad (21c)$$

where $\hbar \equiv h/2\pi$. The causally substantiated version of the differential, 'Dirac quantization' rules is then obtained by using the above definitions of momentum, eq. (4), and energy, eq. (7):

$$p = \frac{\Delta\mathcal{A}}{\Delta x} = -\frac{1}{\Psi}i\hbar\frac{\partial\Psi}{\partial x} , \quad p^2 = -\frac{1}{\Psi}\hbar^2\frac{\partial^2\Psi}{\partial x^2} , \qquad (22)$$

$$E = -\frac{\Delta\mathcal{A}}{\Delta t} = \frac{1}{\Psi}i\hbar\frac{\partial\Psi}{\partial t} , \quad E^2 = -\frac{1}{\Psi}\hbar^2\frac{\partial^2\Psi}{\partial t^2} , \qquad (23)$$

where the wave presentation of higher powers of $p$ and $E$ properly reproduces the wave nature of $\Psi$ [2]. We emphasize the unreduced complex-dynamical meaning of the causally deduced quantization rules: in the basic form of eqs. (21) they present the quantum-beat cycle as a dynamically continuous sequence/entanglement of the virtual soliton 'evolution', in the form of chaotic wave field, during quantum jumps ($\Delta\mathcal{A}$) and its formation during reduction-extension ($\Delta\Psi$). It is therefore natural that quantization rules can be considered as another representation of the same causal definition of space and time (cf. eqs. (4), (7)) emerging within our extended version of the double solution [1,2] and entering now into the delocalised (undulatory) description of complex field-particle dynamics.

Inserting the quantization expressions into corpuscular (Lagrangian) formulation of eqs. (13),

$$E = -L + pv = m_0 c^2 \sqrt{1 - \frac{v^2}{c^2}} + \frac{p^2}{m} , \qquad (13'')$$

we get the dual, delocalised formulation in the form of *wave equation* for the wavefunction $\Psi$ which can be considered as the simplest form of both Klein-Gordon and Dirac wave equations:

$$i\hbar m \frac{\partial \Psi}{\partial t} + \hbar^2 \frac{\partial^2 \Psi}{\partial x^2} - m_0^2 c^2 \Psi = 0 , \qquad (24a)$$

$$-\frac{\hbar^2}{c^2} \frac{\partial^2 \Psi}{\partial t^2} + \hbar^2 \frac{\partial^2 \Psi}{\partial x^2} - m_0^2 c^2 \Psi = 0 . \qquad (24b)$$

$$\frac{\partial^2 \Psi}{\partial t^2} - c^2 \frac{\partial^2 \Psi}{\partial x^2} + \omega_0^2 \Psi = 0 , \qquad (24c)$$

where $\omega_0 \equiv m_0 c^2/\hbar = 2\pi\nu_0$ is the rest-frame 'circular' frequency of the quantum beat (which describes actually the spin vorticity twist [1,2]). More advanced forms of wave equation taking into account e/m interactions with the external field and gravitational effects can be obtained with the help of the same causal quantization procedure [2], providing the causal, complex-dynamical substantiation for the formally identical canonical versions. In the nonrelativistic limit they are reduced to the Schrödinger equation of the canonical form, but automatically provided with the realistic, causally complete interpretation of its structure and solutions.

We can also obtain the Schrödinger equation by directly applying the causal quantization rules, eqs. (22), (23), to the nonrelativistic limit of the energy-momentum relation of eq. (13'') written for the field-particle in the external potential $V(x,t)$ (which causally accounts for the 'expected' dynamic complexity, as it was explained above):

$$E = \frac{p^2}{2m_0} + V(x,t) \quad \rightarrow \quad i\hbar \frac{\partial \Psi}{\partial t} = -\frac{\hbar^2}{2m_0} \frac{\partial^2 \Psi}{\partial x^2} + V(x,t) \Psi(x,t) \qquad (25)$$

(here and above $x$ can be directly extended to the three-dimensional vector of coordinates). The complex-dynamical origin of the Schrödinger equation can also be demonstrated in a transparent form if we rewrite it by multiplying its stationary version (for time-independent $V$) by $(m_0/\hbar^2)\Psi^*(x)$ and integrate over the domain of $\Psi(x)$:

$$q^2 + \frac{m_0}{\hbar} \frac{V_\Psi}{\hbar} = \frac{m_0}{\hbar} \frac{E}{\hbar} , \qquad (26)$$

where

$$V_\Psi = \int_\Omega \Psi^*(x) V(x) \Psi(x) dx, \qquad (27)$$

and

$$q^2 = -\frac{1}{8\pi^2} \int_\Omega \Psi^*(x) \frac{\partial^2 \Psi}{\partial x^2} dx = \frac{m_0}{h} \frac{K}{h}, \qquad (28)$$

with $K$ representing the 'kinetic energy' (below $\Psi(x) \sim \exp(ikx)$):

$$K = -\frac{\hbar^2}{2m_0} \int_\Omega \Psi^*(x) \frac{\partial^2 \Psi}{\partial x^2} dx \sim \frac{\hbar^2 k^2}{2m_0} = \frac{p^2}{2m_0}. \qquad (29)$$

We can see now that the Schrödinger equation, eq. (29), corresponds to the complexity conservation law within the first two sublevels of complexity, expressed in terms of the elementary complexity quantum, $h$. Each number of such quanta within both sublevels is obtained as a product of the internal rest-mass complexity, $m_0/h$, from the lowest sublevel of dynamics by the respective part of the higher-sublevel complexity ($E/h$, $V_\Psi/h$, or $K/h$). Moreover, it is easy to understand that for binding potentials eq. (29) can be satisfied only for those discrete configurations of $\Psi(x)$ that correspond to integral numbers of the same physical complexity quantum (determined by $h$ and corresponding to the quantum-beat cycle, or system 'realisation' change), which explains the famous quantum-mechanical 'energy-level discreteness' by the same universal discreteness (dynamic quantization) of complex behaviour [2].

    The obtained unified complex-dynamical picture of the quantum field mechanics can be explicitly extended to the realistic case of many interacting field-particles. This case corresponds to the general configuration of the world which can be presented as the unique electro-gravitational 'sandwich' consisting from the same two internally continuous protofields/media and producing many randomly emerging field-particles with the described general properties as a result of the protofield interaction [1,2]. In accord with the general hierarchically 'splitting' (or 'branching') character of complex dynamics, there can be several sufficiently stable regimes of the complex-dynamical interaction processes within individual field-particles, differing in their parameters (intensity, spatial scale and configuration), which corresponds to several main types of particles (like leptons, hadrons, and their main subdivisions). The quantum beat process(es) within every field-particle, existing in the *same* couple of the *unique* protofields lead to *interaction* between particles which has two global forms, or transmission channels, corresponding to the two sides, e/m and gravitational, of the 'world sandwich'.

The perturbations of the e/m protofield density produced by the reduction-extension events within each field-particle are at the origin of the e/m interactions between the particles, since the quantum beat parameters (frequency and spatial probability distribution) for each particle depend on the local protofield density and tension. Because of the fundamentally quantized character of quantum beat processes for each field-particle, their mutual influence is exerted also in a *dynamically* quantized character, and the discrete quanta of exchange of e/m protofield perturbations between several particle-processes are none other than the causally extended, realistic version of *photons* and *e/m interaction by exchange of photons*, replacing the corresponding formal prototypes from the single-valued, fundamentally *perturbative* scheme of the canonical 'quantum electrodynamics'. The fundamentally indivisible character of the quantum beat cycle (related also to de Broglie's 'phase accord' concept) provides a causally complete understanding of the origin and quantized character of electric charge and its two 'opposite' varieties. The property of charge emerges as another measure of the same fundamental complexity of the elementary quantum beat process, which provides a physically consistent explanation of the well-known proportionality between the square of the elementary charge, $e$, and Planck's constant: $e^2 = \alpha c \hbar$, where $\alpha \cong 1/137$ is the 'fine structure constant'. The existence of two and only two varieties of electric charge is due to two possible, and opposite, temporal phases of individual quantum beat processes compatible with their universally indivisible character at the lowest level of complexity. In other words, not only all the individual quantum beat processes are periodic in time, but they are temporally coherent (i. e. all the reductions/extensions occur 'in phase'), up to the reversed (opposite) phases for particles with unlike changes (see [2] for more details on the causal interpretation of e/m interactions). Note that another known 'fundamental interaction type', the 'weak' interaction, seems to belong to the same physical channel of interaction transmission through the e/m protofield and can be considered as a different realisation/mode of this channel becoming essential at very short distances, whereas the 'ordinary' e/m interaction is the long-range realisation of the same transmission channel (this assumption is consistent with the experimentally confirmed unification between the e/m and weak interactions, *formally* established within the canonical field theory; see also the end of this section).

The gravitational protofield forms the second universal way of the complex-dynamical 'interaction transfer' through a physically real medium. Namely, the quantum beat process(es) of any field-particle (or group of particles) will inhomogeneously change the 'elastic' properties (effective tension) of the surrounding gravitational medium, and this will influence the frequency and spatial probability distribution of any other field-particle present in thus emerging 'gravitational field' of the first particle(s) (the effect is evidently reciprocal). The binding, particle-forming character of the electro-gravitational coupling shows that this second way of 'indirect' interaction between the field-particles can only result in their *attraction* to each other, irrespective of the charges and forces acting within another, e/m channel of interaction (this shows that the phases of reduction-extension processes are not important here as it is also the case for the property of enertia). This is none other than the mechanism of the *universal gravitation*, possessing all the necessary properties, and simultaneously a causally complete version of the old idea of (gravitational) 'action at a distance' (while its e/m variety is realised at the other side of the 'world sandwich', as described above). The most remarkable feature of this mechanism of gravitational attraction, completing its naturally emerging universality, is its intrinsic unification with the dynamically quantized, dualistic behaviour of the individual field-particle and the inherent causal version of special relativity (time and space relation to motion), which includes the physically based 'equivalence' between gravitational and inertial aspects of mass-energy, the fundamentally *quantum* origin of *any* gravitational effect, and the causal modification of the canonical 'general relativity'. All those results are the direct, physically transparent consequences of the holistic nature of the unreduced complex dynamics (multivalued quantum beat) of one and the same unique 'object', the system of two homogeneously coupled protofields (in particular, the 'strong' fundamental type of interaction can be interpreted, within this picture, as the extreme short-range mode of interaction transmission through the gravitational medium, similar to the 'weak' interaction type with respect to the e/m protofield, see also below). An important additional aspect of this picture of the world, underlying the difference between the two channels of particle interaction, is that its actually existing, observed structure is considerably displaced towards the e/m component (side of the 'sandwich'), whereas gravity manifest itself but indirectly, through effects observed *only* at the e/m side, like universal gravitation and mass (this feature partially explains an aura of mystery and 'irresolvable' problems existing around gravity).

Thus, the inertial aspect of mass-energy is characterised by the (large) total number (per a fixed time period) of the spatially *chaotic* reduction-extensions of the coupled protofields (or virtual soliton jumps) within certain particle or body. It is clear, however, that the more is the number of reductions, observed at the e/m side of the couple in the form of inertia, the stronger is the induced average tension of the surrounding gravitational medium and the stronger will be the induced force of attraction for another particle (or body). In general, both effects, gravitation and inertia, need not be proportional to each other as it happens for moderate, 'nonrelativistic' gravitational interactions (and the stronger, 'relativistic' gravitational effects involve indeed nonlinear dependencies on the inertial masses of the participating bodies [9]). However, what is more important in the resulting holistic picture of the quantum field mechanics, it is that one deals always with the *single complex-dynamical process*, generally with many entangled components, which has its *inseparable* inertial and gravitational aspects/manifestations described above. It is this irreducibly *complex-dynamical* mechanism of 'equivalence' between inertia and gravitation which determines, *due to the underlying chaoticity*, a reasonable *coherence* of the resulting dynamical structure of the world, as it is manifested, for example, in the observed general integrity of the massive cosmic bodies and structures moving in their common gravitational fields. Therefore the generally well-defined, sufficiently inhomogeneous structure of our universe (necessary for any its progressive development) is an indirect, but fundamentally substantiated evidence in favour of the proposed world dynamics uniquely providing the irreducible equivalence (unification) between inertia and gravitation. In a more general meaning of integrity, the proposed sense and mechanism of the equivalence principle provide the unbroken total *cycle* of the world dynamics (in the form of the interaction loop relating the e/m and gravitational protofields) which is a necessary and major 'element' of any quasi-autonomously moving machinery (this particular 'machine of the universe' should be, in addition, self-developing and thus explicitly complex, i. e. dynamically multivalued).

The eventually *quantum origin* of the universal gravitation and any its manifestation (including the extended 'principle of equivalence') is the result of the *complex*, and thus *dynamically* quantized, internal structure of the quantum beat processes within any particle (or their group), the *same one* that provides the previously obtained causal explanation for the wave-particle duality and causal quantization of motion (in this sense, of course, any process or property has a basically 'quantum' character of the same origin, like e. g. any e/m interaction, electric charge, or the inertial mass). Any macroscopically observable gravitational force or effect is but an averaged result of a large number of dynamically quantized 'contractions' of the gravitational medium induced by quantum beat processes within the participating (generally, moving) masses. This internal dynamical quantization of gravity can actually appear in the *explicitly discrete* form of 'quantum gravity' only at extremely small time scales of the order of $\tau_0 = h/(m_0 c^2)$ (~ $10^{-24}$ s for the heaviest particles) and spatial distances of the order of the virtual soliton size (< $10^{-16}$ cm) which actually exist only 'deep within elementary particles' and the related 'exotic' states of matter. Any ordinary 'quantum' scale of motion (in atom and even in the nucleus) is far above this level and therefore all the 'essentially quantum' objects typically behave as 'gravitationally classical'. Nevertheless, the quantum field mechanics naturally incorporates the unique origin and fundamental basis for the explicit 'quantization of gravity' which is impossible within any canonical 'geometric' (formal, non-dynamical) and single-valued (unitary) description of gravity [2].

This is the case, in particular, of the canonical Einsteinian 'general relativity' representing the universal gravitation as a purely geometric 'deformation of (purely abstract, four-dimensional) space-time'. It is clear now that a 'geometrical' presentation of any quantum dynamics, and especially of quantum gravity at the smallest scale, would demand the use of a *dynamically fractal* (and thus also *causally probabilistic*) network of 'geodesics' (or 'curvilinear coordinates'), which can have no sense (any mechanistic simulation within a purely formal, basically single-valued 'noncommutative geometry' cannot either replace a physically based, complex-dynamical approach). As we have seen above, the coherent gravity combination with the (extended) 'quantum mechanics', 'field theory', 'particle physics', and 'special relativity' is possible only when it is presented as an *internally dynamic* (and actually *complex*-dynamical) phenomenon reduced to inhomogeneous distribution of the quantum beat *frequency* due to the corresponding inhomogeneous distribution of *tension* of a *physically real* gravitational medium (protofield, or background). The unitary, necessary single-valued (deterministic) 'geometrisation' can formally be 'successful' only as a *simulation* of the complete complex-dynamical picture in special situations of *externally*, and always *locally*, quasi-regular dynamics, in which case it is analogous to purely *formal* introduction of curvilinear coordinates along 'lines of equal tension' (orthogonal to the 'lines of force') in mathematical description of *integrable* (effectively one-dimensional) dynamics in an inhomogeneous e/m field around fixed charges or elastic-medium deformation around defects (which evidently can be considered as 'deformed space' only in a figurative, non-fundamental sense).

In the unreduced description of gravity the nonuniform tension of the gravitational medium can be associated with the 'gravitational potential', rather than a 'metric'. Thus, for the case of the static gravitational field (cf. [9]) one obtains [2], instead of eq. (2):

$$h\nu_0(x) = m_0 c^2 \sqrt{g_{00}(x)} \ , \tag{30}$$

where the classical metric, $g_{00}(x)$, should be understood rather as an expression containing the gravitational potential (for the case of weak fields, $g_{00}(x) = 1 + 2\phi_g(x)/c^2$, where $\phi_g(x)$ is the classical gravitational field potential [9]). Since $\nu_0(x)$ determines the causal 'time flow' (see above) and $\phi_g(x)$ has the negative sign ($g_{00}(x) < 1$), the above equation substantiates the causal extension of the well-known 'relativistic time retardation' in the gravitational field. This natural involvement of gravity with the dynamically quantized behaviour of micro-objects within the quantum field mechanics can be further specified, including the Dirac wave equation for electron in (relativistic) gravitational and e/m fields [2], but we shall not present here all the details, being interested rather in the fundamental unification of quantum and relativistic behaviour originating from the same unreduced dynamic complexity. The most fundamental, and always present, evidence of this *unique causal origin* of the quantum 'mysteries' and relativistic 'weirdness' is the intrinsically unified mass-energy-complexity determined by the rate (intensity) of the dynamically chaotic quantum beat process and giving simultaneously quantum behaviour, inertia, gravity, and relativistic dynamics (see eqs. (1)-(2), (6)-(9), (13)-(15), (23)-(26), (30)).[3]

Another impasse of the same origin inevitably emerges within the purely 'geometric', clockwork Weltbilt at the scale of the whole universe (i. e. in *cosmology*) because here even any 'ordinary', medium-scale structures play the role of the causally probabilistic, fractal network escaping any feasible presentation within the 'exact-solution' paradigm of the canonical relativity. In this sense the 'curved space-time' solutions of the equations of general relativity for the *whole* universe (like those of Friedmann, and many others) and the related global-dynamics problems (around 'cosmological constant', etc.), can have no causal sense at all, irrespective of the details. All the fractally structured local curvatures of the imaginary 'geodesics' certainly cannot be 'averaged' in a 'global curvature' of the fundamental physical space and time and cannot reproduce the dynamically unpredictable, causally probabilistic evolution of the world. The imaginary curvature of various 'lines of equal tension' can be defined only locally and only for a zero-complexity, effectively one-dimensional behaviour, whereas at the scale of the whole universe the local 'deformations', generally time-dependent and chaotic, can only contribute to the level of *average* tension, or density of the globally 'flat' gravitational medium (this global tension can hardly be properly estimated from within the universe).

The improper, formal mixture of space, being a physically real structure, and time, remaining a non-material sign of the structure emergence, is another characteristic inconsistency of the canonical relativity and its various unitary 'modifications', mechanistically *imitating* the complex-dynamical extension of *dispersion relations* which describe the intensity of space structure emergence in the form of dynamically quantized, probabilistically occurring events. In summary, the causal description of gravity, making an integral part of the quantum field mechanics, gives formally the same results as the canonical general relativity for local, medium-scale, effectively separable problems that actually (and subjectively) have been chosen for its 'experimental verification' (the nonseparable *macroscopic* configurations will necessarily produce the intrinsically chaotic dynamics within our approach, see below). The same 'accord with experiment' can be found for various other approaches to gravity and modifications of the general relativity, which shows that the mentioned 'experimental predictions' are highly 'degenerate' with respect to theoretical schemes applied. On the contrary, the 'degeneracy' is effectively eliminated within critically sensitive understanding of reality in the situations with explicitly complex behaviour (small and big scales, unreduced dynamical chaos), and the unique surviving approach necessarily involves the intrinsic unification of the causally complete extensions of the main fundamental theories and notions.

---

[3]In this connection we should emphasize the fundamental distinction of our results from numerous ambiguous, quasi-philosophical speculations about the involvement of gravity with quantum behaviour, and the canonical 'state-vector reduction' in particular, performed within the ordinary, single-valued approach (see e. g. [13,14] and the references therein). The idea that the *universal*, omnipresent gravitation should somehow influence the quantum behaviour of a massive particle seems to be qualitatively evident. However, it cannot be consistently specified within the unitary science, since the latter, irrespective of the details, cannot propose the irreducible *complex*-dynamical basis for the internal system dynamics which is absolutely indispensable, as we have seen, for the emergent, natural unification of gravity, quantization, and 'special relativity' in the behaviour of an elementary system of *physical* fields with interaction.

The described complex-dynamical unification of quantum (wave) mechanics, relativity, and gravity at the most fundamental level of the world complexity involves, in principle, all the related elementary entities and their properties forming the causally complete picture of reality, and some of them have already been mentioned before. Further analysis within the same concept of complexity [2] permits one to rigorously specify the physically real entities and properties of spin [1,2], magnetic field (self-consistently resulting from the same nonlinear vorticity of the spinning field-particle), electric charge, photon and e/m interaction, the nature of fermions and bosons, their various properties and multi-particle states, etc.

As an important particular result, already mentioned above, note the causal, physically transparent interpretation of the four 'fundamental types of interaction', as well as their natural, dynamical unification within the quantum field mechanics, as opposed to actual failure of all artificial schemes of such 'great unification' within the canonical 'field theory' (even without gravity, remaining practically intractable). Namely, if one recalls that the protofields consist of some very small elements, not directly observable, but most probably not very different in size from the dynamically formed corpuscular state of the elementary field-particle (virtual soliton), then it becomes evident that the 'weak' and 'strong' modes of the unified quantum beat process correspond simply to the localised, short-range interactions between those elements within (and between) the e/m and gravitational protofields, respectively, while the e/m and gravitational interaction types are similarly based on the complementary long-range ('undulatory') transmission of physical influence through the protofield media, with the evident mechanical analogues of those two characteristic modes of behaviour of a quasi-continuous medium. The 'strong' interaction appears thus to be additionally unified with the gravitational one (by the common, gravitational, 'medium of origin' consisting from 'proto-quark' elements), similar to the same type of a closer unification between the e/m and 'weak' interaction types with respect to the e/m protofield (this latter case being formally 'described', but not really understood within the canonical theory). Most important is the fact that all the interaction 'types' are permanently, dynamically mixed within the unique quantum beat process, with its culminating point at the moment of maximal squeeze-reduction realising the profound causal connection between the (complex-dynamical) 'unification of interactions' and the physical origin of the elementary particle as such.[4]

Although we have no place here for further development of these results, the feasibility and the same natural emergence of such non-contradictory unification of the fundamental physical phenomena constitute an important aspect of the intrinsically complete, *extended causality* in physics and science in general, initiated and consistently persuaded by Louis de Broglie starting from his double solution concept. In particular, the universal complexity conservation law [2] mentioned above and expressing the intrinsic physical wholeness of the unreduced complex dynamics can be considered as the direct extension and completion of unification of the basic principles of (i) least action (Maupertuis) from classical mechanics, (ii) shortest optical path (Fermat) from classical optics, and (iii) entropy increase (Carnot, 'second law') from classical thermodynamics, clearly justified by Louis de Broglie and inexcusably missed by the scholar science [7,8,17-19]. The unreduced picture of complex quantum-beat dynamics described above shows that this 'unification of principles' (of the canonical science) is just another aspect of the same intrinsically unified dynamical process that gives also 'unification of entities' (like unification of fundamental forces, elementary particles/fields, and various observed patterns of their behaviour). This kind of reality-based, natural unification should replace its basically insufficient, and therefore inevitably unsuccessful, imitations from the single-valued, unitary science (known as various versions of the 'theory of everything') which try to mechanistically join together the dead pieces of purely formal, one-dimensional description, each of them being irreducibly detached from the living, dynamically complex (multivalued) reality.

---

[4]The fundamental insufficiency of any unitary, mechanistic approach to gravity and its unification with other basic interactions and elements of reality clearly demonstrates the limitations of the single-valued paradigm in general, as they are emphasised especially by its persistent persuasion within 'ultimate unification' ideas in the canonical field theory originating from the Einsteinian program of the 'doubly unitary' field that should give rise to all other fields and entities. The irresolvable difficulty of all those approaches is always reduced to the rigidly fixed, non-developing character of the single-valued, unitary science, where everything new, qualitatively specific can only be artificially inserted 'from outside', in the form of postulates, as it was acutely noticed by Bergson [15] and more specifically by de Broglie [12,16].

## 2. Classical behaviour emergence and the unified hierarchy of world's dynamics

The extended causality should include also dynamic unification with the higher-level systems and behaviour, in the form of their natural emergence from the described level of elementary field-particles. The next higher level of dynamically complex, chaotic behaviour results indeed from the level of the quantum field mechanics in the form of dynamically redundant, and therefore chaotic (complex) behaviour of the *interacting* quantum field-particles at the level of their global, 'averaged' dynamics described above which forms, in its turn, dynamically random sequences of regular de Broglian structures, etc. Dynamic complexity here can take the form of either (Hamiltonian) quantum chaos (in the absence of dissipation) [2,4], or (causally extended) quantum measurement (dissipative, open-system dynamics) [2,3]. The latter can be considered as a complex-dynamical particle interaction process involving formation of a transient quasi-bound state between two (or several) elementary field-particles whose virtual solitons perform, during a transiently short period, their chaotic 'quantum jumps' in a close vicinity of each other, thus temporary reducing the long-range 'wave properties' of the particles. When the (attractive) interaction magnitude is large enough (exceeding the level of the elementary quantum of the global motion energy), such transient bound state of the field-particles can become a stable (permanent) bound state (like that of an atom), and a new *level of complexity* emerges, that of the *classical behaviour*. Indeed, due to the sufficiently strong attraction the bound virtual solitons should always remain close to each other while performing their *random* jumps, which evidently limits the probability of larger 'correlated deviations' to low, exponentially decreasing values and results in a well-localised, trajectorial type of chaotic dynamics of an *isolated* bound system. This naturally emerging classical character of the elementary (and larger) free bound systems is quite different from both canonical *semi-classical* (quasi-classical) behaviour that need not be localised, and canonical (ill-defined) notion of 'classicality', since our elementary classical system need not be macroscopic or dissipative, but is always *internally*, dynamically chaotic (even though macroscopic dissipativity can *quantitatively* amplify the degree of localisation and the ensuing classicality). It becomes clear from the proposed complex-dynamical interpretations of classical behaviour and quantum measurement that in the latter case we deal indeed with a *transiently* emerging, unstable case of classical, localised type of dynamics, which explains the origin of the *postulated* reference to 'classical nature' of the 'instrument of measurement' in the canonical quantum mechanics [20] and shows how exactly the physical manifestations of the *explicitly unified* quantum beat complexity at the lowest levels of being are transferred to higher, classical levels through interaction with the real measurement instruments or any macroscopic bodies which preserves and *amplifies* their diversity, but inevitably destroys their natural unification within the single original process.

    The irreducible role of complex quantum beat dynamics in causally complete interpretation of classical behaviour shows perfect and unique agreement with a number of experimental observations of undulatory, 'quantum' behaviour in various many-particle bound systems *with interaction* that should rather behave as purely classical systems ('bodies') according to practically all canonical interpretations of quantum mechanics invariably evoking (poorly defined) 'decoherence' of a large enough system as the 'explanation' for its expected classical behaviour. However, even systems consisting from a large enough number of elementary particles in order that their 'decoherence' suppress any 'essentially quantum' effect show, for example, unambiguous beam-diffraction effects [21,22] in the canonical type of 'de Broglie wave detection' experiment. We can explain these results, within the above interpretation of classical behaviour, as a transient 'reconstitution' of measurable delocalised, 'undulatory' system properties due to its (non-dissipative) *interaction* with the diffraction slits which leads to 'pulling' of the system (a molecule, a bound cluster of atoms) to one of the slits through the extended state (causal wavefunction) of its unceasing quantum beat process resulting in the 'caterpillar motion' of the system characteristic for the interacting field-particle dynamics (section 1) [2]. Therefore the transient interaction process between the elementary constituents of the diffracting body and the obstacle actually 'develops' the never ceasing internal quantum beat dynamics to an extended spatial scale covering several adjacent slits, which gives the diffraction effects even for this *basically* classical, localised system (in its *free-motion* state) and would be impossible without the underlying reduction-extension cycles of the complex quantum-beat dynamics. Note that the common extension-phase wave field of the system can pull it through a relatively large distance to a slit without problem, even if the total system mass is much greater than that of the elementary constituents contributing to the wavefunction formation, and this is because the mass itself (of the system or any one of its components) *emerges* as a result of the large series of reduction-extension events forming the observed averaged (global) motion (section 1), whereas the internal, detailed system dynamics, giving in particular its transient 'quantum' delocalisation, are driven by the fundamental protofield coupling force which, in terms of energy, is at least of the order of the largest observable elementary particle (or nucleus) rest mass, i. e. 100 GeV [2] (see also e-print gr-qc/9906077). We see thus that the consistently defined classical behaviour of a bound system does not contradict, due to its internal complex-dynamical structure, to a

relative 'amplification' (extension) of its delocalised, undulatory component produced by a suitable, *non-dissipative* system interaction with external objects. These 'quantum', undulatory behaviour of a diffracting classical system will necessarily be limited, however, to the *quasi-classical*, short-wavelength limit of the general quantum behaviour, the case where (always transient) wave delocalization exists, but demands, with the growing system mass, ever higher instrumental resolution for its practical observation and therefore gradually vanishes from observations with the finite-resolution instruments. This explanation agrees with the canonical interpretation of the quasi-classical limit, but provides the nontrivial, qualitative addition, fundamentally lacking to the canonical, single-valued theory, namely the causally explained localised, 'reduced' system behaviour that becomes the 'normal', dominating quality for an elementary bound system in its free state and only partially and transiently changes to a delocalised behaviour under the influence of suitable (non-dissipative) interaction with the environment.

Another generally similar situation of 'quantum behaviour reconstitution' for interacting elementary classical systems is provided by various 'quantum condensates' of atoms and molecules, within a gaseous or liquid 'collective' phase/body: in this case the 'quasi-classical' demands for 'high resolution' appear in the form of (rather severe) limitations from above on the system temperature, or else the necessary internal *spatial* ordering of the quantum beat dynamics for the constituent species will be destroyed [2]. This spatial (non-ideal) 'grating' within 'condensates' is the essential element of their causally complete, complex-dynamical interpretation, absent in the purely abstract canonical approach trying to formally imitate (fit to) only the resulting external 'forms' of averaged behaviour.

The case of quantum measurement interpreted above (see also [2,3]) as a (slightly) *dissipative* interaction of a quantum (including classical) system with the environment (represented by the *a priori* dynamically *regular* 'instrument') is, in a sense, opposite to the described situation of undulatory behaviour of a classical system: due to the dissipative transient binding of several interacting virtual solitons the measured quantum 'object' becomes localised, or classical, for a time period comparable with the characteristic interaction time, so that even an 'essentially quantum', non-classical field-particle (in its *free* or any *non*-dissipative interaction state) demonstrates a momentary suppression of its wave properties that can quickly be reconstituted after the dissipative interaction process is finished. This is what happens, for example, in the canonical field-particle diffraction on slits, if the particle produces an excitation in the slit while passing through it: such particle will not contribute to the interference pattern, but will reconstitute its full undulatory properties almost immediately after passing through the 'dissipative' slit.

It can be seen that these two dualistically opposed situations simply reveal the intrinsic, dynamic duality of the fundamental quantum beat process of protofield interaction reflecting, in its turn, the universal duality between 'compact' system realisations and 'delocalised' transitions between them within any (necessarily) complex interaction dynamics [2] (see also below). It is clear therefore that neither undulatory behaviour of sufficiently large, 'classical' objects, nor irreducible emergence of classicality in any quantum measurement process can ever be consistently explained within the canonical, *single*-valued science incapable to provide any sound basis for the *multiple* (and incompatible) components of dualistic behaviour as such (one should have at least two fundamental components for any dualistic behaviour!). All that remains for the abstract unitary 'interpretations' is to artificially, arbitrarily 'amplify' the ambiguous 'decoherence' effects when they *need* to obtain localisation of an 'essentially quantum' (e. g. 'measured') object or equally arbitrarily 'suppress' them when they *need* to explain the opposite case of 'essentially quantum' behaviour of classically large objects, both kinds of 'rigorous' imitation being achieved by the 'well established' scholar-science method of playing with parameters applied in this case to the vaguely defined 'environment' that can, of course, support any assumptions (contrary to the elementary, common-sense logic).

The hierarchical complexity development continues in the same general fashion to all the higher levels: the quasi-localised trajectory of a classical, bound system becomes dynamically irregular, and thus again delocalised, due to any nontrivial interaction(s) with other classical system(s), then bound, 'condensed' systems of the next levels are formed, etc. It is important that the same qualitative concept of dynamic redundance and the related quantitative description remain valid at any level of (irreducibly complex) reality and are represented by the localised, trajectorial formalism (Hamilton-Lagrange equation generalising the ordinary Hamilton-Jacobi equation) and delocalised, state-functional formalism (generalised Schrödinger equation) [2], the latter extending various canonical equations for quantum and classical 'density matrix' and 'distribution function'. This universality permits us to overcome a persisting general contradiction around (canonical) quantum theory stating that even if its 'peculiar' weirdness could be consistently explained, the world of quantum phenomena would necessarily remain basically separated from the 'ordinary', classical (macroscopic) world, thus preserving at least some of the major contradictions. We see now that the consistent, causal explanation of quantum phenomena is attained *together with* the *truly* consistent, and essentially different from canonical, description of the classical, macroscopic behaviour, within the same concept of the unreduced dynamic complexity. All the 'specific' strangeness of the quantum behaviour is due simply to the fact that it is observed at several *lowest* levels of dynamic complexity, where dissipation (interaction with other levels) is typically absent or occasional and the quite universal process of chaotic realisation change appears in its unveiled, 'explicitly complex-dynamical' (multivalued) form which only *seems* to be weird with respect to unrealistically simplified, single-valued description. Since *any* nontrivial dynamics is complex (intrinsically multivalued), the latter kind of description is *fundamentally* insufficient *also* at all 'classical', *higher* levels of complexity (including all the recent versions of 'dynamic complexity' and 'chaos' developed within the same *single-valued* concept of the canonical science), but at those higher levels an *approximately* or *externally* regular behaviour can be more readily produced (though generally also only in *particular, specially chosen* cases) due to easily accessible, ubiquitous lower-level influences (dissipation, 'control of chaos', etc.). Inside each 'well-defined' pattern (or trajectory) of quasi-regular macroscopic dynamics one can always find many *close*, and therefore practically indistinguishable, but in reality *different* and even *incompatible* among them, component pattern-realisations whose permanent chaotic change is hidden within a somewhat indistinct 'border' of the pattern (or physical width of the trajectory), thus providing another manifestation of the above generalised 'principle of least action' in the form of *real*, causally random (and here relatively small) wandering of the system trajectory/state. This is the regime of generalised 'self-organisation', or 'self-organised criticality' of complex dynamics [2], whereas essentially quantum-mechanical behaviour provides a characteristic example of the complementary regime of 'uniform chaos', where all the realisations are sufficiently and *equally* different, so that their (properly frequent) change cannot remain unnoticed and leaves an impression of a mysteriously 'many-faced', 'ill-defined' and chaotically 'flickering' reality (this is also the case of many *explicitly* chaotic *macroscopic* phenomena).[5] Therefore one always deals with the same, universal hierarchy of complex, redundantly multivalued dynamics which simply turns to us one or another of its infinitely diverse 'realisations', aspects or regimes, depending on the level of complexity and observation conditions.

---

[5]One can recall here the well-known objections of Einstein to the standard quantum mechanics insisting on the necessity to find well-defined 'real states' for any physical system which, according to him, should necessarily be totally *deterministic*. As an example of such kind of realism, he evokes the position of the centre of mass of the Moon and states that nobody can deny that this position is always absolutely and objectively precise [23]. We know now why the latter statement is practically wrong and cannot be true in principle: a system of many interacting bodies has a nonseparable, chaotic dynamics, and the position of the centre of mass of each of them for any given moment can be defined only approximately (although for weakly interacting massive bodies, like the Sun and planets, indeterminacy is indeed relatively small, it is always finite, even without any additional, 'noisy' perturbations). This example clearly demonstrates the fundamental difference between the Einsteinian *mechanistic* realism which is impossible without the basic, eventual *total determinism*, ("God does not play dice") and the realism of causal randomness which does not reject indeterminacy, but instead demystifies it by clearly showing its *purely dynamic* origin and proposing well specified means for its calculation. This difference manifests itself in the difference between the corresponding two ways of unified description of the world: Einstein was looking for a unified description of a clock-work, predictable world, a kind of the 'ultimate exact solution' giving all the others in the respective limits, whereas the unification of the unreduced science of complexity gives an always *unpredictably* changing, developing world provided, however, with the objectively determined, *predictable probability* of any its event.

It would be not out of place to conclude this presentation of the extended version of de Broglie's double solution with the obtained causally complete interpretation of the 'canonical', and at the same time never properly understood, basic expression for the de Broglie wavelength, eq. (9). Taking into account the above unifying picture of the underlying quantum beat dynamics, we can now present this remarkably compact expression of the unreduced dynamic complexity of the field-particle, $h = \lambda_\text{B} m v$, in the form of a qualitative, physical, but rigorously substantiated relation between the emerging, dynamically connected aspects of complexity:

$$\left\{\begin{matrix} \boldsymbol{h} \\ \textit{quantized} \\ \textit{complexity} \end{matrix}\right\} = \left\{\begin{matrix} \boldsymbol{\lambda_\text{B}} \\ \textit{wave behaviour,} \\ \textit{undulatory space structure} \end{matrix}\right\} \otimes \left\{\begin{matrix} \boldsymbol{m} \\ \textit{relativistic inertial and} \\ \textit{gravitational mass-energy} \end{matrix}\right\} \otimes \left\{\begin{matrix} \boldsymbol{v} \\ \textit{corpuscular behaviour,} \\ \textit{trajectorial space structure} \end{matrix}\right\}.$$

The fundamental involvement of dynamic complexity in the causally complete understanding of quantum dynamics was prodigiously predicted, not only in general, but also in many essential details, by the author of the above relation who wrote [16]:[6]

> "Il me paraît plus naturel et plus conforme aux idées qui ont toujours heureusement orienté la recherche scientifique de supposer que les transitions quantiques pourront un jour être interprétées, peut-être à l'aide de moyens analytiques dont nous ne disposons pas encore, comme des processus très rapides, mais en principe descriptibles en termes d'espace et de temps, analogues à ces passages brusques d'un cycle limite à un autre que l'on rencontre très fréquemment dans l'étude des phénomènes mécaniques et électromagnétiques non linéaires."

---

[6]English translation (A.K.): "It seems to me more natural and more consistent with the ideas that always fortunately oriented scientific research to suppose that one will be able, one day, to interpret quantum transitions, probably with the help of analytical means that are not yet at our disposition, as very rapid processes which nonetheless can, in principle, be described in terms of space and time and are analogous to those sudden passages from one limit cycle to another which are very frequently met within the studies of nonlinear mechanical and electromagnetic phenomena."